\documentstyle[twocolumn,prl,aps,floats]{revtex}
\begin{document}
\preprint{BROWN-HET-1139, hep-ph/9808471}
\draft
\def\la{\langle }
\def\ra{ \rangle }
 \def\bea{\begin{eqnarray}}
\def\eea{\end{eqnarray}}
 \def\gmf{\gamma _{5}}
\renewcommand{\topfraction}{0.99}
\renewcommand{\bottomfraction}{0.99}
\twocolumn[\hsize\textwidth\columnwidth\hsize\csname 
@twocolumnfalse\endcsname
  
\title
{\Large Baryogenesis with QCD Domain Walls}
       
\author{R. Brandenberger$^{1,2}$, I. Halperin$^2$ and 
A. Zhitnitsky$^2$}

\address{~\\$^1$Department of Physics, Brown University, 
Providence, RI 02912, USA;
~\\$^2$Department of Physics and Astronomy, University of 
British Columbia,
Vancouver, BC, V6T 1Z1, CANADA} 
\maketitle
\begin{abstract}
We propose a new baryosymmetric mechanism for baryogenesis which 
takes place at the QCD scale and is based on the existence 
of domain walls separating the 
metastable vacua from the lowest energy vacuum. The
walls acquire fractional negative and positive baryon charges,
while the observed baryon asymmetry is due to a non-zero value of 
the $ \theta $ angle at the temperatures near the QCD 
chiral phase transition.
The regions of metastable vacuum bounded by walls carry a negative 
baryon charge and  
may contribute a significant fraction of the dark matter of 
the Universe. 
  
\end{abstract}

]

\vskip 0.4cm
{\bf 1.} The origin of the asymmetry between baryons and antibaryons, 
and, more specifically, the origin of the observed baryon to
 entropy ratio $n_B / s \sim 10^{-9}$ ($n_B$ being the net 
baryon number density, and $s$ the entropy density) is still 
a mystery. In order to explain this number from symmetric 
initial conditions in the early Universe, it is generally 
assumed 
that three criteria, first laid down by Sakharov \cite{Sakharov} 
must be satified:
(1) 
there must exist baryon number violating 
processes;  
(2) 
these 
processes must involve $C $ and $ CP $ violation; 
and (3)
they must take place out of 
thermal 
equilibrium.

Many mechanisms for baryogenesis have been proposed over the past 
twenty years (for recent reviews see e.g.
\cite{Dolgov,ShapRub,Trodden}), 
but all involve new physics, often at extremely high energy 
scales. A 
potential problem for all baryogenesis mechanisms which operate 
at 
scales 
higher than the electroweak scale $\Lambda_{\rm{EW}}$ is that 
electroweak sphaleron effects wash out any pre-existing baryon 
number unless other 
quantum numbers such as $B - L$ (where $B$ and $L$ are baryon and 
lepton number, respectively) are violated during baryogenesis. 
Therefore, in recent years electroweak baryogenesis has received a 
lot of attention, see e.g. \cite{ShapRub,Trodden}. In the standard 
model, however, $ CP$ 
violation is too small to generate the observed value of $n_B / s$. 
In 
addition, given the present bounds on the Higgs mass, it appears 
unlikely that in the standard model the electroweak phase 
transition 
is sufficiently strongly first order to generate the  
out-of-equilibrium bubbles required to satisfy the third 
Sakharov criterium. This problem can be circumvented within 
topological defect-mediated electroweak 
baryogenesis \cite{BD,BDPT},
but only at the cost of introducing new physics above the
electroweak scale.

Recently, it was suggested \cite{HZ1} that the Sakharov criteria
could be satisfied at the QCD scale.
In this paper, we propose an explicit   mechanism which
can  explain the observed value of $n_B /s$ and 
which takes place at the QCD scale without    introduction of any 
new physics. 
Our mechanism, however, is strictly 
speaking not a baryogenesis, but rather a charge separation
 mechanism, since the total baryon current is exactly conserved. 
In this aspect, our approach makes use of 
ideas proposed in
 the context of toy models in \cite{Dodelson,Dolgov}.

The first essential ingredient in our scenario is 
the realization that there exist several vacua in 
QCD whose energy degeneracy is broken only by a very small 
amount $ \Delta E \sim m_q $ ($m_q $ is a quark mass) 
compared to the heights of 
the barriers between the vacua. This  
leads to
the domain walls which provide the locus of 
out-of-thermal equilibrium required by the third Sakharov 
criterion.

This picture of the vacuum structure is based on the 
analysis of the general properties of the $ \theta $ dependence 
in 
QCD. It is known that 
 physical values depend on the strong $ CP $
parameter $\theta$ through $\theta/N_f$ (or $ \theta/N_c $ in
pure gluodynamics,
$N_f $ and $ N_c $ are the number of flavors and colors, 
respectively)
and, at the same time, they
must be $ 2 \pi $ periodic functions of $\theta$. 
This is only possible if an effective potential is  
multi-valued, and in the partition function one
sums over all branches \cite{KS,HZ2}.
 This yields the vacuum structure
described above. 
In terms of the original theory 
this prescription means summation over all
topological classes $
Z\sim\sum_n \int DA 
\exp \left[ -S_0+iQ(\theta+2\pi n)  
  \right]$
 imposing  the topological charge $ Q $ quantization. 
This picture arises
in supersymmetric \cite{KS} and  
non-supersymmetric \cite{HZ2} theories. The appearance of domain 
walls in gauge theories was also advocated by Witten within 
different
frameworks \cite{Witten}. 
  
The second crucial input is the fact that the domain walls 
may carry 
a nonvanishing fermion number, a 
well-known \cite{JR,GW,NS} property 
of many solitonic configurations. The domain walls are also 
locations 
of maximal $CP$-violation \cite{HZ1} due
to the phase difference of the chiral 
condensate across the wall, demonstrating that the second 
Sakharov criterium is satisfied. In what follows  
we use a specific 
low energy effective Lagrangian for QCD \cite{HZ2}, 
though one can expect that the 
general properties described above are inherent in  
QCD, and should appear in any approach.

Making use of these ingredients in the context of the early 
Universe, 
it follows by the Kibble-Zurek mechanism \cite{Kibble,Zurek} 
that at 
the QCD phase transition at the temperature $T_c \sim 200 \, 
{\rm MeV}$ 
a network of 
domain walls separating different vacua will form.  
Because of the energy bias $ \Delta E $, walls 
surrounding regions of metastable vacua will tend to 
contract. However,
when coupled to fermions, 
they can be stabilized, yielding  
some compact objects, 
not necessarily of spherical form,
to be called {\it B-shells} in what follows.
(This name is due to the fact that the  
(anti-) baryon charge is concentrated on the surface
of such a compact object. This bears some resemblence to  
the SLAC bag \cite{SLAC}.) 
The compensating positive baryon 
number is left behind in the bulk. The {\it B-shells}, in 
turn, can 
contribute a substantial fraction of the dark matter of the 
Universe.
 
{\bf 2.}
 We start by describing the effective potential \cite{HZ2}
which allows one to analyze the vacuum properties and domain walls 
in QCD. 
In this approach, the Goldstone fields are 
described by the unitary matrix $ U_{ij} $ corresponding to the
$ \gmf $ phases of the chiral condensate: 
$ \la \bar{\Psi}_{L}^{i} 
\Psi_{R}^{j} \ra 
=  - | \la \bar{\Psi}_{L} \Psi_{R} \ra | \, U_{ij} $.
  The  effective chiral potential is periodic 
in $ \theta $ and takes the form  
\bea
\label{11} 
W_{QCD}(\theta,U)=-\lim_{V \rightarrow 
\infty}  \frac{1}{V} \log  \{ 
 \sum_{l=0}^{p-1} \exp  [ 
V E \cos\Phi \nonumber \\ 
 + \frac{1}{2} V \, Tr ( M U + M^{+} U^{+} )  ]\} \; ,  
\eea
 where $V$ is the 4-volume and 
\[
\Phi=- \frac{q}{p} \theta + i\frac{q}{p} \log  Det \, U  
+ \frac{2 \pi}{p}l \; , \;  
M = diag (m_{i}  | \la \bar{\Psi}^{i} \Psi^{i} 
\ra | )
\]
($ m_i $ are the quark masses),
while 
$ E = \la b \alpha_s /(32 \pi) G^2 \ra $ is the QCD gluon 
condensate.
Two integers $ p $ and $ q $ 
are related to a discrete
symmetry which is a remnant of the anomaly and can be found 
only by 
explicit
dynamical calculations. In the large $N_c$ limit, $q/p\sim 
1/N_c$
in order for  the $U(1)$ problem to be solved.
The number of metastable vacua is very sensitive to the 
values of $p, q$. In an approach of \cite{HZ2,FHZ} $ p =3b = 
11N_c - 
2 N_f , q = 8 $ ($N_f $ is the 
number of flavors), but here we keep them as free 
parameters. 
  
One can argue \cite{HZ2,FHZ} that Eq. (\ref{11}) represents the 
anomalous effective Lagrangian realizing broken conformal and 
chiral symmetries of QCD. The arguments are the following: 
(a) Eq. (\ref{11}) correctly reproduces
 the Di Veccia-Veneziano-Witten
 effective chiral Lagrangian \cite{Wit2}
in the large $ N_c $ limit; (b) it reproduces the 
anomalous 
conformal and chiral Ward identities of QCD; (c) it contains the
built-in topological charge quantization, which shows up
in (\ref{11}) via the sum over the integers $ l $ and 
presence of cusp singularities at certain values  of the fields.
 
Thus, the effective chiral potential (\ref{11}) satisfies all 
general requirements on the theory and leads to the above 
picture of the vacuum structure and the related domain walls.
For the domain walls interpolating between the 
true vacuum of lowest energy and the first excited state
at $ \theta = 0 $,
the surface energy density $ \sigma $ of the wall  
 for the potential (\ref{11}) has been calculated in  
\cite{HZ1,FHZ}:
\begin{equation} 
\label{tension}
\sigma \, = \, \frac{4 p}{q \sqrt{N_f} }  
f_{\pi}\sqrt{E}
\, \left( 1 - \cos \frac{\pi}{ 2 p} \right) +0(m_q 
f_{\pi}^2 ) \; ,
\end{equation}
where $ f_{\pi} = 133 \; MeV $.
The energy splitting between the ground state and the 
metastable 
state at $\theta = 0$ is 
\begin{equation}
\label{10}
 \Delta E \, =  \, m_q N_f \left| \la \bar{\Psi}\Psi\ra \right| 
\left( 1-\cos  \frac{2\pi}{qN_{f}} \right)  +0(m_q^2) \, .
\end{equation}
Due to this energy difference, regions of the metastable state
tend to disappear, unless they are stabilized
by coupling to fermions. In this case, one expects to 
end up with blobs of the metastable state (B-shells).

{\bf 3.}
Next we would like to argue that the above domain walls may 
acquire a fractional baryon charge when coupled to the baryons.
As
our domain wall configuration is a flavor singlet (the $ \eta' $
domain wall \cite{FHZ}), 
we keep only the
relevant part of the matrix  $U = \exp [i\alpha(z)] $,
where $\alpha(z)$ has a solitonic shape
written explicitly in\cite{FHZ}. Neglecting the isospin, we 
consider the following simplified Lagrangian  
for the nucleon $ N $ 
interacting with the external, non-fluctuating chiral
field  $ U $ and additionally via a four-fermion interaction:
 \bea
 \label{lag4}
{\cal L}_4 \, = \, \bar{N} i\partial_{\mu}\gamma_{\mu}N - 
m_{N} \bar{N}_{L} U N_{R} - m_{N} \bar{N}_{R} U^{+} N_{L}  \\
- \lambda \left( \bar{N}_{L} N_{R} \right) \left( 
\bar{N}_{R} N_{L} \right) \nonumber \; 
\eea
(the choice of the sign of $ \lambda >0 $ corresponds to the 
repulsion in the U(1) channel). 

The important point is that the nucleon mass $ m_{N} $ takes 
different values in different vacua: $ m_{N} = m_{N}^{(0)} + 
m_q f( \theta , k ) $ , where $ m_{N}^{(0)} $ is the nucleon mass in 
the chiral limit and $ k $ labels different vacua. This is related 
to the fact that the chiral condensate varies by an amount 
$ \sim m_q $
in different vacua. Given this, estimates of a function $  
f( \theta , k ) $ are possible but will not be considered here. 
Thus, in the adiabatic approximation to the interaction with the 
domain
wall, $ m_{N} $ should be considered as a slowly varying function
of $ z $. Fortunately, its precise 
form is irrelevant for our purposes, only the asymptotia enter 
the final
answer, see Eq. (\ref{b2}) below.

Because of the planar symmetry, the problem of computing the 
charge $B^{(4)}=\int\bar{N}\gamma_0 N d^3x$
in our four dimensional (4D) theory (\ref{lag4}) reduces 
in the mean field approximation to the 
calculation of the corresponding 
charge $B^{(2)}=\int\bar{\psi}\gamma_0\psi dz$
 of a two dimensional  (2D) theory. Omitting the details of 
this calculation which will be given elsewhere, we here present the 
final answer for the 2D baryon number on the wall:    
\begin{equation}
\label{b2}
B^{(2)} = \left. \frac{1}{\pi} \, \arccos \frac{m_{N}}{ 
 \tilde{\lambda} } \right|_{z= - \infty}^{z = + \infty} \; , 
\end{equation}
where $ \tilde{\lambda} $ has dimension of mass and can be found 
in terms of 
$ \lambda, m_{N}^{(0)} $ and a degeneracy factor $ n = N/S $, 
see below.

To find the original 4D baryon charge $ B^{(4)} $, we should 
take account of a degeneracy related  to the symmetry under 
the  shifts  along the wall plane. 
In many body physics
the definition of the charge is $B=\int\sum_i\bar{N^i}
\gamma_0 N^i d^3x$
where we should sum over all particles
(possible quantum  states). In our case such a summation 
yields
\bea
\label{1}
Q \equiv B^{(4)}=B^{(2)}g\int\frac{dxdydp_xdp_y}{
(2\pi)^2}\equiv 
B^{(2)}N \; ,
\eea
where $g=2\cdot 2$ describes the degeneracy in spin and isospin.
For a fixed number of quantum states 
$N=gS\int\frac{d^2p}{(2\pi)^2}
=\frac{gSp_F^2}{4\pi}$ ($S$ is the area of 
the wall), the fermi energy $\bar{E}_F$
 of the domain-wall fermions is determined  by 
\bea
\label{2}
\bar{E}_F= gS\int\frac{pd^2p}{(2\pi)^2}=\frac{2}{3}Np_F=
\frac{4}{3} \, \sqrt{\frac{\pi}{g}}\frac{N^{3/2}}{\sqrt{S}}.
\eea
The size of the surface which can accommodate the fixed number 
of fermions $N$ can be found from the minimization equation
$ \frac{d \bar{E}_{0}}{ dS}|_{N=const}=0$,  where  
the total energy of the fermions
residing on a surface $S$ is given by 
\bea
\label{3}
\bar{E_0}= \sigma S+
\frac{4}{3} \, \sqrt{\frac{\pi}{g}}\frac{N^{3/2}}{\sqrt{S}} \; .
\eea
The minimization then
relates the density of fermions
per unit area $n= N/S$ to the 
tension $\sigma$ (\ref{tension}). We obtain 
\bea
\label{6}
Q= - ST_c^2\alpha_1, ~~\alpha_1=  
- \frac{\sigma^{2/3}}{T_c^2}(\frac{9g}{4\pi})^{1/3} B^{(2)} \; .
\eea

{\bf 4.}
Now we are ready 
to introduce our proposed baryogenesis (charge 
separation) mechanism. At the QCD phase 
transition at the temperature $T_c \simeq
200 \, MeV $, the chiral 
condensate forms. Because of the presence 
of nearly degenerate states, a network of domain walls will 
arise immediately after 
$ T_c $. At $ \theta = 0 $, there are two degenerate
metastable states $ |B \ra $ and $ |C \ra $ above 
the true vacuum of lowest energy $ |A \ra $. 
For simplicity, we ignore metastable 
states of higher energy, although they might be important 
for evolution of the domain wall network \cite{Vil}. 
A $ CP $ transformation exchanges the states $ |B \ra $ and 
$ | C \ra $.
Respectively, the baryon charge 
of a  $ A-B $ wall will be opposite to that of a  $ A-C $ wall, and 
no baryon asymmetry can be produced in this case. The situation 
becomes 
different, however, when $ \theta \neq 0 $ at the temperatures 
close to 
$ T_c $. This case will be considered below. (It is implied 
that the strong $ CP$  problem will be cured by an axion at lower 
temperatures. At  
$ T \sim T_c $, the axion field 
does not yet settle in its 
ground state, and thus $ \theta (T_c) $ might be of order unity.
Note that as long as an initial value $ \theta(T_c) $ is the same 
in the entire observable Universe, so will be the sign of the baryon
asymmetry. This occurs if the Universe undergoes inflation either 
after or during the Peccei-Quinn symmetry breaking.)  
In this case there is a splitting 
$ \Delta E \sim \theta m_q $
between the energies of lowest metastable vacua, which 
translates into 
the splitting $ \Delta M \sim \theta M $ for the masses of B-shells 
with negative and positive baryon charges (here $ M $ stands for the 
B-shell mass at $ \theta = 0 $). If the B-shells 
reach thermal equilibrium at a
temperature $ T $, 
the relative densities of negative and positive B-shells will 
be fixed
by the Boltzmann formula, and one type of B-shells
will thus be very strongly suppressed. 
(A similar interplay between
the spontaneous and explicit breaking of $ CP $  was considered
in \cite{AW} within a different scheme.)  
Assuming that this is the case, we will therefore 
discuss the evolution of the 
domain walls of only one (negative) type.

The initial wall separation depends on the details of the
chiral phase transition. We will assume  that an infinite domain 
wall network will exist until a temperature $ T_d $, 
after which it decays into a number of finite clusters of the 
false vacuum.  The typical
wall separation $ \xi = \xi(T_d) $ at this temperature depends on 
the initial wall separation at formation, the details of the damping 
mechanism and the interplay between the energy bias (\ref{10})
and the surface pressure in the walls \cite{Kibble,Zurek,walls}.
For the time being, we will use the value of  $ \xi = \xi(T_d) $
as an unknown parameter in our scenario, and relegate a 
brief discussion of possible values of $ \xi $ to the 
end of the paper.
  
Given the typical wall separation $ \xi $,
the total area in the walls 
at this temperature is \cite{walls} $ \la S \ra 
\simeq V/ \xi $, where 
$ V $ stands for the Hubble volume at $ T_d $. Note that this 
is parametrically larger by $ V^{1/3} / \xi $ than the Hubble 
area $ V^{2/3} $. This enhancement
means that the baryon charge on the wall
is in fact not a surface effect, as Eq. (\ref{6}) would naively 
suggest, but rather a volume effect. 
Thus, the total (anti-) baryon charge stored
in the (negative) walls at $ T_d $ is fixed by  
Eq. (\ref{6}), while  
the compensating baryon charge is left in the bulk (the
positive walls and their subsequent decay
products are effectively ascribed to the bulk).  
As the entropy density is 
$s = g_* T_d^3$, where $g_*\sim 10$
 is the number of spin degrees of freedom in the radiation bath 
at $T_d$, we may estimate the baryon to entropy ratio in the bulk 
at $ T_d $:
\begin{equation}
\label{td}
\left. \frac{n_B}{s} \right|_{T_d} \simeq \frac{ \alpha_1  
 T_{c}^2}{ g_*  \xi  T_{d}^3 } \, . 
\end{equation}

After $T_d$, the domain wall network will break up into 
finite clusters of the false vacuum 
(``B-shells"). 
Although the  question of stability of these B-shells 
requires a detailed
study, qualitatively  
we expect that the bubbles will shrink, but 
not decay completely since they will eventually be 
stabilized by the fermions.
As the   
non-relativistic baryons can hardly cross the wall, we expect 
the shells to be stable against the escape of baryons from the
interior, but able to lose heat by baryon pair annihilation
and emission of the 
photons and/or neutrinos. The quantum stability
of the B-shells 
will be addressed in \cite{BHZ2}. Generally, one 
expects an exponential suppression of quantum decays 
by the baryon charge of the surface. 

To proceed, we introduce two dimensionless  
constants $ \alpha_{2} $ and $ \alpha_3 $ by parametrizing the 
total area and volume of the B-shells as $ \alpha_{2}^2 V/ \xi $ and 
$ \alpha_3 V $, respectively. (This parametrization does not 
imply any specific assumption about the  form of the B-shells.)  
In addition, we  neglect
the change of the Hubble radius between the time corresponding to 
$ T_d $ and the time at which the B-shells are formed.  
We thus obtain 
\begin{equation} 
\label{barent}
{{n_B} \over s} \, \simeq \, \alpha_1 \alpha_2^2 g_*^{-1} 
 \frac{ T_{c}^2}{ \xi T_{d}^3 } = \alpha_{2}^2 \; 
\left. \frac{n_B}{s} \right|_{T_d} \; .
\end{equation}
The energy density $\rho_B$ in the blobs of the metastable state 
will redshift as matter, i.e. $\rho_B(T) \sim T^{3}$. Hence, the 
B-shells can contribute to the dark matter of the Universe.  
Their contribution to
\begin{equation} 
\Omega_B \, = \, {{\rho_B(t_{eq})} \over {\rho_r(t_{eq})}} 
= \frac{ \rho_{B}(t_d)}{ g_* T_{d}^3 T_{eq}} \, ,
\end{equation}
(where $t_{eq}$ is the time of equal matter and radiation, and 
$\rho_r$ 
is the energy density in radiation which, until $t_{eq}$, 
dominates 
the total energy density) is, roughly, 
\begin{equation} 
\label{omegab}
\Omega_B \, \simeq \, \alpha_3 \, \frac{2 \pi^2}{q^2 N_f} \,
\frac{ m_q | \la \bar{\Psi} \Psi \ra | }{ g_* T_{d}^3 T_{eq}} \, ,
\end{equation} 
where we used the expansion of the cosine in Eq. (\ref{10}). 
This estimate 
implies that the dominant contribution to the energy density
$ \rho_{B} $  
inside the B-shells is that due to the false vacuum energy 
(\ref{10}),
which is larger than the non-vacuum contribution $ \sim  \alpha_3 
m_{N} n_{B} $ with $ n_{B} $
given by (\ref{barent}). 
(Here we assume that the  
baryon density $n_B $ is of the same order of 
magnitude in the bulk and interior,
while the redundant baryons and anti-baryons with zero net $ n_B $
will eventually annihilate and leave the B-shells together with 
a flux of photons and neutrinos.)

Comparing (\ref{barent}) and (\ref{omegab}), we see that
the resulting baryon asymmetry and contribution of the 
B-shells to the dark matter density are related with each other 
via the geometrical parameters $ \alpha_2 , \alpha_3 $ of the 
distribution of shells sizes. To calculate them is a 
complicated task which is at the moment beyond our ability. 
However, it is of interest to reverse the argument and estimate 
 the parameters $ \alpha_2 , \alpha_3 $ assuming that 
the B-shells do provide the observed baryon asymmetry $ n_{B}/s \sim
10^{-9} $ and contribute significantly to the dark matter, 
$ \Omega_B \sim 1 $. As can be seen from (\ref{barent}) and 
(\ref{omegab}), this requires $ \alpha_{3} \sim 10^{-7} $ and   
$ \xi T_d \simeq 10^{6} 
\alpha_{2}^2 $,  if $ \alpha_1 \sim 10^{-3} $ 
and $ T_d \sim T_c $.
As $ \alpha_2 < 1 $ and $ \xi T_d  \gg 1 $ for the 
thin wall approximation 
to work, this leaves us with the window $ T_{c}^{-1} 
\ll \xi < 10^{6} T_{c}^{-1} $ for the proposed 
mechanism to be operative. This corridor is consistent with 
the Kibble-Zurek scenario for the topological defects formation
in the early Universe \cite{Kibble,Zurek}. 

In conclusion, qualitative as our arguments are, they suggest that 
baryogenesis can proceed at the QCD scale, and might be tightly
connected with the origin of the dark matter in the Universe.
An alternative scenario could also be considered, in which the charge 
separation proceeds due to different interactions of particles and 
anti-particles with a CP-breaking domain wall with zero baryon charge, 
similarly to what happens in electroweak baryogenesis. The blobs of 
metastable vacua might be stabilized in this case by the 
fermi pressure
of anti-baryons  trapped inside during the evolution of the walls.
We emphasise that the suggested ideas can be, in principle,
experimentally tested at RHIC. A more quantitative and detailed 
analysis will be given in \cite{BHZ2}.

This work was supported in part by the Canadian NSERC 
and by the U.S. Department of Energy under 
Contract DE-FG02-91ER40688, TASK A. We wish to thank D. Schwarz,
E. Shuryak, A. Vilenkin  
and L. Yaffe for 
valuable comments.

\end{document}